# SWAPHI: Smith-Waterman Protein Database Search on Xeon Phi Coprocessors

Yongchao Liu, Bertil Schmidt
Institut für Informatik
Johannes Gutenberg Universität Mainz
Mainz, Germany
Emails: {liuy; bertil.schmidt}@uni-mainz.de

*Abstract*—The maximal sensitivity of the Smith-Waterman (SW) algorithm has enabled its wide use in biological sequence database search. Unfortunately, the high sensitivity comes at the expense of quadratic time complexity, which makes the algorithm computationally demanding for big databases. In this paper, we present SWAPHI, the first parallelized algorithm employing Xeon Phi coprocessors to accelerate SW protein database search. SWAPHI is designed based on the scale-and-vectorize approach, i.e. it boosts alignment speed by effectively utilizing both the coarse-grained parallelism from the many co-processing cores (scale) and the fine-grained parallelism from the 512-bit wide single instruction, multiple data (SIMD) vectors within each core (vectorize). By searching against the large UniProtKB/TrEMBL protein database, SWAPHI achieves a performance of up to 58.8 billion cell updates per second (GCUPS) on one coprocessor and up to 228.4 GCUPS on four coprocessors. Furthermore, it demonstrates good parallel scalability on varying number of coprocessors, and is also superior to both SWIPE on 16 high-end CPU cores and BLAST+ on 8 cores when using four coprocessors, with the maximum speedup of 1.52 and 1.86, respectively. SWAPHI is written in C++ language (with a set of SIMD intrinsics), and is freely available at http://swaphi.sourceforge.net.

## I. INTRODUCTION

The Smith-Waterman (SW) [1] [2] algorithm is a critical and fundamental operation in many biological applications such as biological database search [3] [4], multiple sequence alignment [5] [6] and next generation sequencing read alignment [7-10], due to its maximal sensitivity in identifying optimal local alignments. This algorithm is able to compute the optimal local alignment score of a pair of given sequences in linear space, but has a quadratic time complexity in terms of sequence length. The quadratic runtime makes the SW algorithm computationally demanding for big sequence databases, and has therefore motivated a substantial amount of research to reduce the runtime through parallelization on high-performance computing architectures such as clusters/clouds [11] and accelerators [12]. Recent research has mainly focused on the use of the accelerators, including field programmable gate arrays (FPGAs), single instruction multiple data (SIMD) vector processing units (VPUs) on CPUs, multi-core Cell Broadband Engine (Cell/BE), and general-purpose graphics processing units (GPUs), especially compute unified device architecture (CUDA)-enabled GPUs.

For FPGAs, linear systolic arrays [13] [14] and custom instructions [15] have been proposed to effectively accelerate the SW algorithm. For SIMD VPUs on CPUs, two general methods have been investigated. One is the inter-sequence (or inter-task) parallelization model, which performs multiple alignments in individual SIMD vectors with one vector lane computing one alignment (e.g. [16] [17]). The other is the intra-sequence (or intra-task) parallelization model, which computes in parallel the alignment of a single sequence pair in the SIMD vectors based on two computational patterns: SIMD computation parallel to minor diagonals in the alignment matrix [18] and SIMD computation parallel to the query sequence by means of a sequential [19] or striped [20] layout. Up to date, the intra-sequence model has attracted more research efforts than the inter-sequence model. However, compared to the intra-sequence model, the major advantages of the inter-sequence model are the independent alignment computation in SIMD vectors as well as the runtime independence of scoring schemes used. These two models offer a general computational framework for other accelerators with SIMD VPUs such as Cell/BEs and GPUs. On Cell/BEs, few SW implementations have been proposed [21] [22] and all of them are designed based on the intra-sequence model. On GPUs, initially open graphics library (OpenGL) was used to program the SW algorithm [23]. As the advent of the CUDA programming model, a number of implementations [24-31] have been developed using CUDA, among which the CUDASW++ software package [24] is popular.

In this paper, we present SWAPHI (Smith-Waterman Algorithm on Xeon PHI coprocessors), a parallelized algorithm which for the first time has employed Xeon Phi coprocessors to accelerate the SW protein database search. This algorithm gains high speed by coupling both the coarse-grained parallelism from the many processor cores on each coprocessor and the fine-grained parallelism from the 512-bit wide SIMD vectors within each processor core. Performance evaluations, by searching against the UniProtKB/TrEMBL (TrEMBL) protein database, show that SWAPHI achieves a performance of up to 58.8 GCUPS on a single coprocessor and up to 228.4 GCUPS on four coprocessors. In addition, on the TrEMBL database SWAPHI has demonstrated good parallel scalability for varying number of coprocessors sharing the same host.

## II. BACKGROUND

### A. Smith-Waterman algorithm

Given a sequence $S$, we define $S[i, j]$ to denote the substring which starts at position $i$ and ends at position $j$, and $S[i]$ to denote the $i$-th residue. For a sequence pair $S_1$ and $S_2$, the recurrence of the SW algorithm with affine gap penalties is defined as

$$
\begin{aligned}
H_{i,j} &= \max\{0, H_{i-1,j-1} + fsbt(S_1[i], S_2[j]), E_{i,j}, F_{i,j}\} \\
E_{i,j} &= \max\{E_{i-1,j} - \alpha, H_{i-1,j} - \beta\} \\
F_{i,j} &= \max\{F_{i,j-1} - \alpha, H_{i,j-1} - \beta\}
\end{aligned} \quad (1)
$$

where $H_{i,j}$, $E_{i,j}$ and $F_{i,j}$ represent the local alignment score of prefixes $S_1[1,i]$ and $S_2[1,j]$ with $S_1[i]$ aligned to $S_2[j]$, $S_1[i]$ aligned to a gap and $S_2[j]$ aligned to a gap, respectively. $\alpha$ is the gap extension penalty, $\beta$ is the sum of gap open and extension penalties, and $fsbt$ is a scoring function, usually represented as a scoring matrix, which defines the matching and mismatching scores between residues. The recurrence is initialized as $H_{i,0} = H_{0,j} = E_{0,j} = F_{i,0} = 0$ for $0 \leq i \leq |S_1|$ and $0 \leq j \leq |S_2|$. The optimal local alignment score is the maximal alignment score in the alignment matrix $H$ and can be calculated in linear space.

### B. Xeon Phi coprocessor

A Xeon Phi coprocessor is a many-core shared-memory architecture running a specialized Linux operating system (OS) [32]. This coprocessor consists of a set of (around 60) processor cores and provides full cache coherency over the entire chip. Each core offers four-way simultaneous multi-threading, enabling a coprocessor to have up to 4×number-of-cores concurrent threads at full capacity. However, for a specific application, the number of concurrent threads per core should generally be a tunable parameter.

While offering scalar processing, each core also includes a newly-designed VPU which features a 512-bit wide SIMD instruction set architecture (ISA). In each VPU, there are 32 512-bit vector registers, each of which can be split to either 16 32-bit-wide lanes or 8 64-bit-wide lanes. Unfortunately, the coprocessor does not provide support for legacy SIMD ISAs such as the streaming SIMD extensions (SSE) series. This means that applications based on such SIMD instructions cannot be compiled and executed on the coprocessor, but require re-writing with the new SIMD instructions in order to exploit the compute power of the coprocessor. As for the caching mechanism, the coprocessor configures each core to have a 32 KB L1 instruction cache (I-cache), a 32 KB L1 data cache (D-cache) and a 512 KB L2 cache. Moreover, all L2 caches across the entire system are interconnected, with each other as well as the memory controllers, by means of a bidirectional ring bus. This interconnect efficiently creates a shared L2 cache of over 30 MB across all cores. Fig. 1 gives a high-level overview of the coprocessor.

There are two usage models for invoking the coprocessor: offload model and native model. The offload model relies on language extensions for offload (LEO), represented as compiler directives, to offload highly-parallel parts of an application to the coprocessor. This model sends input data and code to the coprocessor at startup time of an offload region, and then transfers back the output data to the host when the offload computation completes. The LEO is a compiler-assisted solution to allow developers to tag offload regions for execution on the coprocessor. As mentioned above, the coprocessor is a Linux-based many-core computer. This means that the code inside any offload region theoretically can contain any code as well as routine calls, and can use a number of parallel programming models including *OpenMP* and *Pthreads* [33] [34]. The native model treats a coprocessor as a normal symmetric multi-processing computer with Linux. This model cross-compiles a program and executes it natively on the coprocessor by logging into the system. Compared to the offload model, the native model avoids the overhead incurred by offloading for initialization, data transfer and kernel invocation, but is not flexible to collaborate with the host and other coprocessors. In this paper, we have therefore investigated the offload model to coordinate multiple coprocessors to perform sequence alignments.

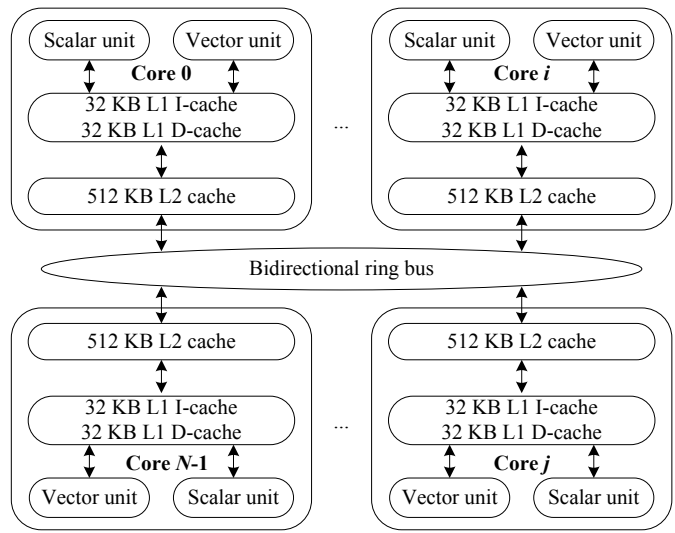

Fig. 1 High-level overview of the Xeon Phi coprocessor

## III. SCALE-AND-VECTORIZE PARALLELIZATIONS

SWPAHI is based on the "scale-and-vectorize" approach, i.e. it gains high speed by utilizing coarse-grained parallelism from multi-threading across all processor cores as well as fine-grained parallelism from the 512-bit wide SIMD vectors within each core. Our algorithm has investigated two SIMD vectorized implementations based on the inter-sequence model and the intra-sequence model, respectively. On the coprocessor, any 512-bit SIMD vector is split to 16 lanes with each lane occupying 32 bits. Compared to the conventional 128-bit SSE vectors that can also be split to 16 lanes (but with 8-bit lane width), this split does not make our parallelization have more vector lanes, but enables us to avoid the redundant computations for the alignments with indicative score overflows.

Both implementations share the same program workflow (see Fig. 2). It works in four stages: (*i*) construct a query profile for the query if applicable, depending on the alignment configurations; (*ii*) perform alignments by creating as many host threads as the number of coprocessors used, where one host thread corresponds to exactly one coprocessor; (*iii*) wait for the completion of all host threads; and (*iv*) sort all alignment scores in descending order and output the alignment results. In Stage (*ii*), each host thread carries out and coordinates the offloading operations to its corresponding coprocessor. To alleviate memory pressure on the coprocessor, each host thread loads the database sequences onto the coprocessor chunk-by-chunk at runtime. To support big databases and achieve good load balance, we build indices for the input database offline prior to alignment and store the index files on disk. All subject sequences are sorted in ascending order of sequence length for the indexing. The index files have been carefully organized so that they can be mapped into virtual memory and directly accessed as normal physical memory.

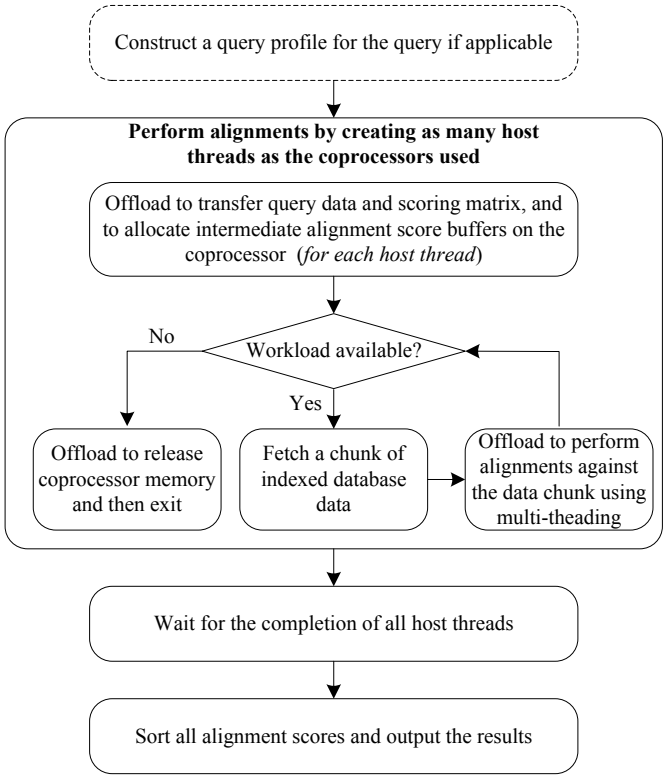

Fig. 2 Program workflow of SWAPHI

### A. Multi-threading on the Coprocessor

We have employed the OpenMP parallel programming model to implement multi-threading on the coprocessor and have configured each coprocessor to have a maximum of 240 concurrent device threads (default setting and configurable by users). For the host thread, after obtaining a chunk of database sequences from its pool of workloads (see Fig. 2), the host thread reaches the offload region, invokes the corresponding coprocessor to execute the code in the offload region, and then waits until the offloaded computation completes. On the coprocessor, the core loop, which performs the alignments between the query and the chunk of database sequences, is parallelized using OpenMP directives. In this loop, each device thread carries out one alignment at a time. For the parallelized loop, four kinds of loop scheduling polices, namely *auto*, *static*, *dynamic* and *guided*, can be specified to control the distribution of loop iterations over all device threads. Through our evaluations, the *static* scheduling performs worst. This is caused by the irregular computation among all loop iterations due to the varying lengths of subject sequences. As for the other three scheduling policies, the *guided* scheduling outperforms the others more frequently, albeit by a slight margin. Based on this observation, we have selected the *guided* scheduling as default.

For the linear-space implementation of the SW algorithm, intermediate buffers are required to store one row of matrices $H$ and $E$ (in our case), respectively. Considering that the sizes of the intermediate buffers are directly proportional to the query length, we have pre-allocated them for each device thread before offloading to conduct alignments. This pre-allocation enables us to avoid frequent memory allocations and de-allocations on the coprocessor. For a certain query, each device thread will re-use its own intermediate buffers until the completion of all alignments. To make faster SIMD computation with 512-bit vectors, we have carefully managed the memory allocations so that the addresses of the intermediate buffers for each device thread are 64-byte-aligned, which is the size of a SIMD vector as well as the cache line-size of the L1 and L2 caches.

In addition, within each device thread, we have employed the fine-grained parallelism from SIMD instructions, and used a set of Intel C++ compiler intrinsic functions to implement the SW alignment kernels (see Table 1).

### B. Inter-sequence SIMD parallelization

*1) Sequence profile:* To facilitate 16-lane SIMD computing, we create a two-dimensional sequence profile [31] of size 16×$L$ for any non-overlapping 16 consecutive subject sequences in the database. $L$ is the maximum length of the 16 sequences and must be a multiple of 8. Each sequence in the sequence profile has been properly padded with dummy residues, whose substitution scores for any residue is zero, to satisfy the constraint on $L$. Each row of a sequence profile forms a 16-lane residue vector with each lane occupying 8 bits and can be loaded to any 512-bit 16-lane vector *vecDB* using the *_mm512_extload_epi32* intrinsic function (see Fig. 3). Our inter-sequence model considers a sequence profile as a unit to build database indices as well as distribute workloads over all coprocessors. Moreover, each device thread is assigned to process the alignment of one sequence profile at a time.

TABLE I.     INTEL C++ COMPILER INTRINSIC FUNCTIONS USED IN SWAPHI

| Category | Intrinsic functions | SIMD parallelization models | |
|---|---|---|---|
| | | Inter-sequence | Intra-sequence |
| Vector mask | _mm512_int2mask | | ✓ |
| Arithmetic | _mm512_add_epi32 | ✓ | ✓ |
| | _mm512_mask_sub_epi32 | ✓ | |
| Compare | _mm512_cmpge_epi32_mask | ✓ | |
| | _mm512_cmpgt_epi32_mask | | ✓ |
| Initialization | mm512_set_epi32 | ✓ | ✓ |
| | mm512_setzero_epi32 | ✓ | ✓ |
| Maximum | _mm512_max_epi32 | ✓ | ✓ |
| Load | _mm512_load_epi32 | ✓ | ✓ |
| | _mm512_extload_epi32 | ✓ | ✓ |
| Shuffle | _mm512_permutevar_epi32 | ✓ | |
| | _mm512_mask_permutevar_epi32 | ✓ | ✓ |
| Store | mm512_store_epi32 | ✓ | ✓ |
| | _mm512_packstorelo_epi32 | | ✓ |
| | _mm512_packstorehi_epi32 | | ✓ |

*2) Query profile:* Based on the inter-sequence model, we have investigated two representations of substitution scores: sequential-layout query profile [19] and score profile [17], to tune the performance of our algorithm. Given a query $Q$ defined over an alphabet $\Sigma$, we represent a sequential-layout query profile as a two-dimensional array of size $|Q|\times|\Sigma|$. Each column $r$ of the query profile is comprised of substitution scores required for aligning the whole query to the residue $r\in\Sigma$. Since $|\Sigma|$ is less than 32 for biological sequences, each row of the query profile is extended to contain 32 elements for faster data loading from memory to vector registers. Given a residue vector register *vecDB* and a query position, we can realize the substitution scores between them by a single gather operation (intrinsic *_mm512_i32extgather_epi32*) with *vecDB* as indices. However, we found that this gather operation is not as lightweight as expected, albeit the good data locality of the accessed query profile row. Thus, we have conceived a more efficient approach to extract substitution scores from a query profile. Fig. 3 shows the code segments used for substitution score loading from a query profile.

*3) Score profile:* A score profile is a two-dimensional array of size $|\Sigma|\times N$, which defines the substitution scores between residue $r\in\Sigma$ and $N$ successive residue vectors of a sequence profile. Compared to a query profile, a score profile usually has a smaller memory footprint and more convenient vector loading of substitution scores. However, the drawback is that we must reconstruct the score profile for every $N$ residue vectors in the sequence profile. At times this extra computational overhead may offset the earned speed for relatively short queries. Furthermore, $N$ should be tuned for better performance based on the characteristics of the underlying hardware. In our inter-sequence model, $N$ is set to 8. The score profile can be constructed from the scoring matrix and the sequence profile. To achieve fast construction, similar to the query profile, we have extended each row of the scoring matrix to contain 32 elements with each element taking 8 bits. Example code segments for the score profile construction are shown in Fig. 4, where $N$=2.

### C. Intra-sequence SIMD parallelization

Unlike the inter-sequence model that treats a sequence profile as a unit, our intra-sequence model considers an individual subject sequence as a unit to build database indices as well as distribute workloads. In this model, each device thread receives one subject sequence at a time and performs parallel SIMD alignment between the query and the subject sequence based on the striped approach [20]. This approach employs a striped-layout query profile to realize fast vector loading of substitution scores (with the *_mm512_extload_epi32* intrinsic function). As there are 16 lanes in a SIMD vector, we require the query length to be multiples of 16 in order to construct the striped query profile (otherwise padding the query with dummy residues). In addition, several technical issues, including saturation arithmetic operations, shift operations and predicate operations, must be addressed for the 512-bit SIMD vectors.

Saturation additions/subtractions are required to ensure that alignment scores are non-negative and do not overflow. Since each SIMD vector lane occupies 32 bits in our implementation, we merely need to ensure that all scores are always non-negative. In this case, we have used the maximum instruction (the *_mm512_max_epi32* intrinsic function) to mimic the required saturation operations. As for both shift and predicate operations, the coprocessor offers direct hardware support. A shift operation can be realized by means of the shuffle instructions (the *_mm512_mask_permutevar_epi32* intrinsic function), and a predicate operation by means of the compare instructions (intrinsic *_mm512_cmpgt_epi32_mask*). Table 1 lists the intrinsic functions that have been used in the intra-sequence model.

**(a) Load and pre-process subject sequence residues (outer loop of the SW algorithm)**
vecInt16 = _mm512_set_epi32(16, 16, 16, 16, 16, 16, 16, 16, 16, 16, 16, 16, 16, 16, 16, 16); /*offset register*/
/*load one residue vector from the subject sequence profile (__m128i* __restrict__ sequences)*/
vecDB = _mm512_extload_epi32(sequences, _MM_UPCONV_EPI32_UINT8, _MM_BROADCAST32_NONE, 0);
/*compare each residue index with 16 and returns a vector mask*/
vecMask = _mm512_cmpge_epi32_mask(vecDB, vecInt16);
/*adjust the residue indices that are greater than or equal to 16*/
vecDB = _mm512_mask_sub_epi32(vecDB, vecMask, vecDB, vecInt16);
**(b) Load substitution scores for each query position (inner loop of the SW algorithm)**
/*load the low and high 16 elements of the query profile row (__m128i * __restrict__ qprfRow)*/
vecLo = _mm512_extload_epi32(qprfRow, _MM_UPCONV_EPI32_SINT8, _MM_BROADCAST32_NONE, 0);
vecHi = _mm512_extload_epi32(qprfRow + 1, _MM_UPCONV_EPI32_SINT8, _MM_BROADCAST32_NONE, 0);
/*get the substitution scores*/
vecSubScore = _mm512_permutevar_epi32(vecDB, vecLo);
vecSubScore = _mm512_mask_permutevar_epi32(vecSubScore, vecMask, vecDB, vecHi);

Fig. 3 Code segments used for substitution loading from a query profile

```
vecInt16 = _mm512_set_epi32(16, 16, 16, 16, 16, 16, 16, 16, 16, 16, 16, 16, 16, 16, 16, 16); /*offset register*/
/*load residue vectors from the subject sequence profile (__m128i* __restrict__ sequences)*/
vecDB = _mm512_extload_epi32(sequences, _MM_UPCONV_EPI32_UINT8, _MM_BROADCAST32_NONE, 0);
vecMask = _mm512_cmpge_epi32_mask(vecDB, vecInt16);
vecDB = _mm512_mask_sub_epi32(vecDB, vecMask, vecDB, vecInt16);
vecDB2 = _mm512_extload_epi32(sequences + 1, _MM_UPCONV_EPI32_UINT8, _MM_BROADCAST32_NONE, 0);
vecMask2 = _mm512_cmpge_epi32_mask(vecDB2, vecInt16);
vecDB2 = _mm512_mask_sub_epi32(vecDB2, vecMask2, vecDB2, vecInt16);
for (r = 0; r < |∑|; r++){
  /*load the low and high 16 elements of the scoring matrix row r (__m128i* matrix; __m512i* scorePrf)*/
  vecLo = _mm512_extload_epi32(matrix++, _MM_UPCONV_EPI32_SINT8, _MM_BROADCAST32_NONE, 0);
  vecHi = _mm512_extload_epi32(matrix++, _MM_UPCONV_EPI32_SINT8, _MM_BROADCAST32_NONE, 0);
  /*get the substitution scores and store them*/
  vecSubScore = _mm512_permutevar_epi32(vecDB, vecLo);
  vecSubScore = _mm512_mask_permutevar_epi32(vecSubScore, vecMask, vecDB, vecHi);
  _mm512_store_epi32(scorePrf++, vecSubScore);
  /*get the substitution scores and store them*/
  vecSubScore = _mm512_permutevar_epi32(vecDB2, vecLo);
  vecSubScore = _mm512_mask_permutevar_epi32(vecSubScore, vecMask2, vecDB2, vecHi);
  _mm512_store_epi32(scorePrf++, vecSubScore);
}
```

Fig. 4 Code segments used for the score profile construction with $N=2$

## IV. PERFORMANCE EVALUATION

### A. Comparison of SWAPHI variants

We have first evaluated the performance (in terms of GCUPS) of the three different variants of our algorithm: inter-sequence model with score profile (InterSP), inter-sequence model with query profile (InterQP) and intra-sequence model with query profile (IntraQP), by searching 20 query protein sequences against the TrEMBL protein database (release 2013_08). The TrEMBL database comprises 13,208,986,710 amino acids in 41,451,118 sequences, with the longest sequence containing 36,805 amino acids. All queries have lengths ranging from 144 to 5,478, and are publicly available in the UniProtKB/Swiss-Prot database. Their accession numbers are listed as follows in ascending order of sequence length: P02232, P05013, P14942, P07327, P01008, P03435, P42357, P21177, Q38941, P27895, P07756, P04775, P19096, P28167, P0C6B8, P20930, P08519, Q7TMA5, P33450, and Q9UKN1. We have conducted all tests on a compute node with two Intel E5-2670 8-core 2.60GHz CPUs and 64 GB memory running the Linux OS. This node is further equipped with four Intel Xeon Phi coprocessors (B1PRQ-5110P/5120D), each of which has 60 active processor cores (at a clock frequency of 1.05 GHz) and 7.9 GB device memory. In this evaluation, all variants have used the same scoring scheme, i.e. scoring matrix BLOSUM62 and a gap penalty of $10-2k$.

Fig. 5 illustrates the performance of all variants for varying query lengths. On a single coprocessor, the average and maximum performance is 54.4 and 58.8 GCUPS for InterSP, 51.8 and 53.8 GCUPS for InterQP, and 32.8 and 45.6 GCUPS for IntraSP, respectively. On four coprocessors, the average and maximum performance goes up to 200.4 and 228.4 GCUPS for InterSP, 191.2 and 209.0 GCUPS for InterQP, and 123.3 and 164.9 GCUPS for IntraQP, respectively. The performance of both InterSP and InterQP consistently improves for increasing query lengths. InterSP outperforms InterQP for the queries of lengths ≥ 375, whereas the latter performs better than the former for all others. This can be explained by the additional overhead incurred by the score profile construction, which cannot be effectively offset by the alignment computation for shorter queries. The intra-sequence variant does not show consistently increasing performance as the query length grows, but shows some fluctuations. However, a positive observation is that the variant reaches the maximum performance at queries of lengths around 464, which is very close to the average sequence length of 318 in the TrEMBL database. By comparing the implementations based on the two models, we can see that the two variants based on the inter-sequence model demonstrate superior performance to the variant based on the intra-sequence model. In consideration of the runtime sensitivity to the scoring schemes, we conclude that the inter-sequence model is more advantageous for searching big biological database such as TrEMBL.

Furthermore, we have evaluated the parallel scalability of all variants in terms of the number of coprocessors (see Fig. 6). We can observe that each of the three variants shows good scalability on two and four coprocessors. On two coprocessors, InterSP achieves an average speedup of 1.95 and a maximum speedup of 2.00; InterQP yields an average speedup of 1.95 and a maximum speedup of 1.99; and IntraQP produces an average speedup of 1.97 and a maximum speedup of 2.03. On four coprocessors, the average and maximum speedup is 3.66 and 3.90 for InterSP, 3.68 and 3.89 for InterQP, and 3.78 and 4.04 for IntraQP, respectively.

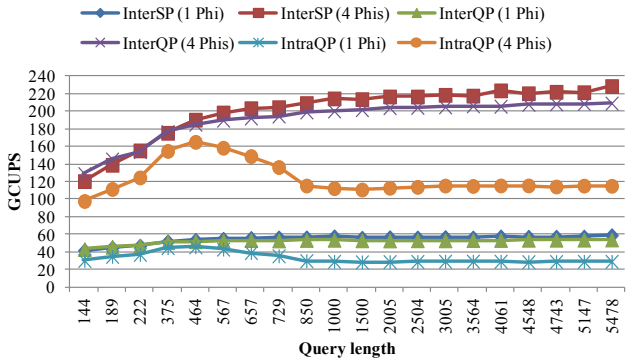

Fig. 5 Performance comparison between the three variants of SWAPHI

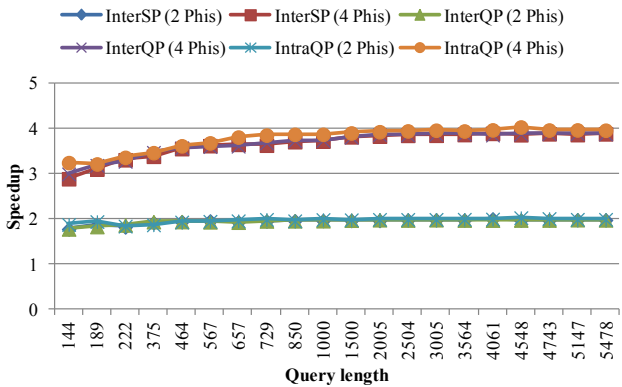

Fig. 6 Scalability of different variants in terms of number of coprocessors

## B. Comparison to SWIPE and BLAST+

Secondly, we have compared our algorithm to the CPU-based counterparts: SWIPE (v2.0.7) [17] and BLAST+ (v2.2.28) [4] by searching against the TrEMBL database. In this evaluation, SWAPHI used the variant InterSP and all evaluated algorithms used the default scoring schemes. Considering that the runtime of BLAST+ is sensitive to scoring schemes, we have attempted to run it with other scoring matrices such as BLOSUM50 and PAM250, but failed as a result of exceptional execution errors. Therefore, we have merely included the performance of BLAST+ with scoring matrix BLOSUM62 and a gap penalty of 11-1$k$ (default settings) in this evaluation. In addition, other parameters "–b 0 –v 0" and "–num_alignments 0" have been used for SWIPE and BLAST+, respectively. All tests have been carried out on the aforementioned compute node.

Fig. 7 shows the performance comparisons to both SWIPE and BLAST+. On 8 CPU cores, SWIPE achieves an average performance of 80.1 GCUPS with a maximum of 84.0 GCUPS, while BLAST+ yields an average performance of 174.7 GCUPS with a maximum of 272.9 GCUPS. On 16 CPU cores, the average performance and the maximum performance go up to 149.1 and 157.4 GCUPS for SWIPE, and 318.6 and 498.4 GCUPS for BLAST+, respectively. For each query, SWAPHI on four coprocessors could not outperform BLAST+ on 16 CPU cores, but is superior to SWIPE on 16 CPU cores. Compared to BLAST+ on 8 cores, SWAPHI performs better for most queries and runs 1.19× faster on average (1.86× maximally). Compared to SWIPE on 8 and 16 cores, SWAPHI gives a speedup of 2.49 and 1.34 on average (2.83 and 1.52 maximally), respectively.

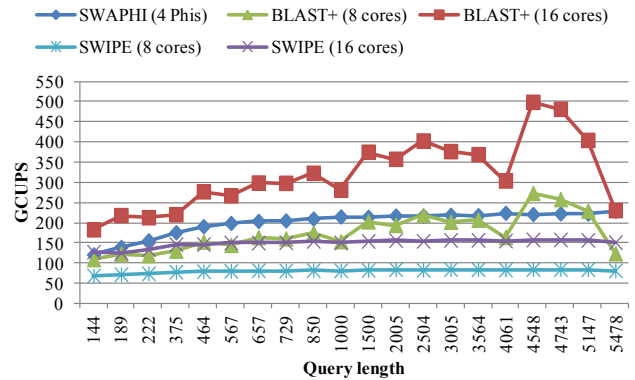

Fig. 7 Performance comparison to SWIPE and BLAST+

## C. Comparison to CUDASW++ 3.0

Finally, we have compared our algorithm to CUDASW++ 3.0 (v3.0.14) [31], which is currently the fastest SW implementation on CUDA-enabled GPUs for protein database search. Because CUDASW++ 3.0 does not support databases as big as TrEMBL, we have utilized the UniProtKB/Swiss-Prot database (release 2013_08: 192,091,492 amino acids in 540,732 sequences) instead. Moreover, CUDASW++ 3.0 employs a hybrid CPU-GPU parallelism by distributing workloads among both CPUs and GPUs. For fair comparisons, we have merely evaluated the performance of the GPU-only version of CUDASW++ 3.0 by disabling CPU threads. However, the GPU-only version does not support subject sequences of lengths >3072 (by default), and we have therefore created a new reduced Swiss-Prot database by extracting all subject sequences of lengths ≤3072 from the original Swiss-Prot database. This new database comprises 99.88% sequences and 98.43% amino acids of the original one. The performance of the GPU-only CUDASW++ 3.0 has been evaluated on a single GeForce GTX Titan (Titan) graphics card, which is designed based on the Kepler architecture. This Titan GPU contains 2,688 scalar processor cores in 14 streaming multiprocessors, has 6 GB device memory and runs at a clock rate of 875.5 MHz. In this evaluation, for all queries SWAPHI used the variant InterSP, while CUDASW++ 3.0 used the query profile variant.

Fig. 8 shows the performance comparisons between SWAPHI and CUDASW++ 3.0 on the reduced Swiss-Prot database. On the Titan GPU, the GPU-only CUDASW++ 3.0 yields an average performance of 108.9 GCUPS and a maximum performance of 115.4 GCUPS. However, the maximum outcome of our algorithm is merely 53.2, 90.8 and 124.6 GCUPS on one, two and four coprocessors, respectively, although the performance keeps growing as the query length becomes larger. On a single coprocessor, the performance of SWAPHI on this reduced database is relatively consistent with that on the TrEMBL database. However, when using multiple coprocessors, the performance is inferior to that on the TrEMBL database, especially in the case of four coprocessors. This is because the small workload assigned to each coprocessor could not spur sufficient computations to offset the additional runtime overhead incurred by the offloading. However, based on the evaluation using the TrEMBL database, we can reasonably infer that for big databases, the performance of our algorithm on two coprocessors is supposed to be comparable to that of CUDASW++ 3.0 on a single Titan GPU.

coprocessor (by means of multi-threading) and the fine-grained parallelism from SIMD vectors within each processor core (by means of vectorized computation). Moreover, SWAPHI provides support for multiple coprocessors sharing the same host in order to further boost the performance. By searching against the big TrEMBL database, our algorithm yields a performance of up to 58.8 GCUPS on a single coprocessor and up to 228.4 GCUPS on four coprocessors. In addition, SWAPHI demonstrates good parallel scalability in terms of number of coprocessors.

We have further compared our algorithm to the top-performing CPU-based and GPU-based counterparts: SWIPE, BLAST+ and CUDASW++ 3.0. On the TrEMBL database, SWAPHI on four coprocessors could not perform better than BLAST+ on 16 CPU cores for all queries, but achieves an average speedup of 1.19 over the latter on 8 CPU cores. Compared to SWIPE, our algorithm demonstrates superior performance and runs 2.49× and 1.34× faster than the former on 8 and 16 CPU cores, respectively. The comparison to CUDASW++ 3.0 was carried out by searching against a reduce Swiss-Prot database, as the GPU-only CUDASW++ 3.0 does not support big database and merely supports subject sequences of lengths ≤3072 (by default). The performance evaluation shows that on a single coprocessor our algorithm is inferior to the GPU-only CUDASW++ 3.0 on a single high-end Titan graphics card. However, based on the performance of SWAPHI on the big TrEMBL database, we can reasonably infer than the performance of our algorithm on two coprocessors is comparable to that of the GPU-only CUDASW++ 3.0 on a Titan GPU.

Based on our programming and evaluation experiences, we have observed some computational characteristics of Xeon Phi coprocessors. Firstly, device memory accesses on the coprocessor are still heavy in some sense, albeit with two-level caching and higher memory bandwidth. We have therefore adopted a tiled SW computation, as did in some GPU-based implementations (e.g. [24] [25]), in order to significantly reduce the number of memory accesses to the intermediate buffers. Secondly, data accesses should be aligned as much as possible. The coprocessor requires an alignment to 64 bytes and is able to realize aligned data allocation by means of compiler storage-class attributes or specialized dynamic memory allocation functions e.g. _mm_malloc(). Thirdly, the gather intrinsic functions are not as lightweight as expected, even if the data accesses has good locality. In the current implementation, we have only exploited Xeon Phi coprocessors to accelerate the SW protein database search. An approach to further boosting the performance can be the concurrent execution of alignments on both CPUs and coprocessors by means of a hybrid parallelism model. This hybrid model has been shown to be effective by some GPU-accelerated applications (e.g. [31], [35-37]). In addition, fast backtracking of optimal local alignments on coprocessors is also an important research problem and can be considered as part of our future work.

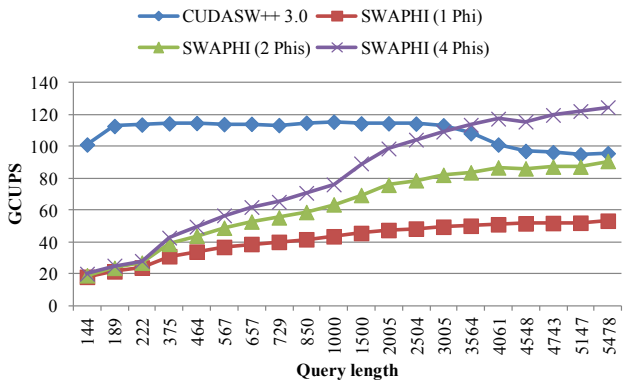

Fig. 8 Performance comparison to CUDASW++ 3.0 on Swiss-Prot database

## V. CONCLUSIONS

We have presented SWAPHI, the first SW protein database search algorithm on Xeon Phi coprocessors. This algorithm achieves high performance by effectively coupling the coarse-grained parallelism from the many processor cores on the

ACKNOWLEDGMENT

We thank Tim Süß and Markus Tacke for their help to set up the compute node with four Xeon Phi coprocessors.